\documentclass
[aps,prl,twocolumn,floatfix,english,showpacs,10pt]{revtex4-2}%
\usepackage{graphicx}
\usepackage{booktabs}
\usepackage{array}
\usepackage{makecell}
\usepackage{amsmath}
\usepackage{physics}
\usepackage{amssymb}
\usepackage{colordvi}
\usepackage{verbatim}
\usepackage{xcolor}
\usepackage{mathrsfs}
\usepackage{epsfig}
\usepackage{lipsum}
\usepackage{amsfonts}
\usepackage{makecell}
\usepackage{esint}
\usepackage{bm}
\usepackage[most]{tcolorbox}

\usepackage[unicode=true, breaklinks=false, pdfborder={0 0 1}, backref=false,
colorlinks=true, linkcolor=blue, urlcolor=blue, citecolor=blue]{hyperref}%
\providecommand{\U}[1]{\protect\rule{.1in}{.1in}}
\providecommand{\U}[1]{\protect\rule{.1in}{.1in}}
\setcitestyle{numbers,square}

\begin{document}

\title{Frequency Comb of Electric-Polarization Waves}

\author{Xiyin Ye}
\affiliation{School of Physics, Huazhong University of Science and Technology, Wuhan 430074, China}

\author{Tao Yu}
\email{taoyuphy@hust.edu.cn}
\affiliation{School of Physics, Huazhong University of Science and Technology, Wuhan 430074, China}

\date{\today }

\begin{abstract}
Frequency combs are a spectrum of equally spaced frequency components with very high time–frequency accuracy, which have been widely used in the optical and microwave frequency ranges. We propose the realization of a frequency comb operating at the terahertz regime in terms of the nonlinear dynamics of electric-polarization waves, or ferrons as their quanta, in the ferroelectric materials. The efficiency of the frequency comb of the electric-polarization waves is exactly proportional to the static electric polarization carried by the ferron modes, which thereby offers new opportunities for the direct observation and application of the intrinsic properties of ferrons. 
\end{abstract}

\maketitle

\textit{Introduction}.---An optical frequency comb consists of equally spaced and phase-coherent spectral components. Originating from mode-locked laser research~\cite{optical_comb1,optical_comb9}, it has revolutionized techniques such as precision measurement, optical atomic clocks, astronomical spectrography, and quantum information~\cite{optical_comb2,optical_comb3,optical_comb4,optical_comb5,optical_comb6,optical_comb7,optical_comb8}. 
Recently, the frequency comb concept has been successfully extended to other quasiparticle systems, such as phonons~\cite{phonon1,phonon2,phonon3,phonon4,phonon5,phonon6,phonon7,phonon8,phonon9,phonon10} and magnons~\cite{magnon1,magnon2,magnon3,magnon4,magnon5,magnon6,magnon7,magnon8,magnon9}. However, acoustic phonon and magnon frequency combs typically operate at microwave or lower frequencies, while the terahertz regime remains challenging for conventional comb technologies.

Recently, several theoretical studies have predicted the existence of electric-polarization fluctuation waves in ferroelectrics~\cite{Tang2022,Tang2024,Bauer2021,Tang2022_thermoelectric,Hu2022,Rodriguez-Suarez2024,Zhou2023,Zhu2024,Morozovska2025}, with their quanta known as ferrons~\cite{Bauer2022,Bauer2023,Yu2026_review}. Analogous to magnons in magnets that can carry a spin angular momentum~\cite{Kranendonk1958,Stancil_book,Chumak2015}, the key characteristic of ferrons, distinguishing from phonons, is their ability to carry a large \textit{static} electric polarization~\cite{Tang2022_thermoelectric,Bauer2021}. Predictions such as thermal transport of electric polarization~\cite{Tang2022,Bauer2023}, directional/hyperbolic routing of electric polarization~\cite{Zhou2023}, nonlocal thermoelectric effects~\cite{Tang2023}, and strong coupling to superconductors~\cite{Nursagatov2026} demonstrate the potential of ferrons in future technology. Experimentally, coherent laser~\cite{Choe2025,Zhang2025} or terahertz~\cite{Subedi2026} excitation, nonlocal electrical injection and detection~\cite{Shen2025}, and electric-field control of thermal transport~\cite{Wooten2023} reveal already several interesting optical and transport properties related to ferrons. While these measurements provide evidence for ferron-mediated phenomena, they probe macroscopic responses rather than the static electric dipole moment carried by ferrons, i.e., the key characteristic of ferron. Direct tomography of this static dipole moment is essential for revealing the direct evidence and microscopic nature of ferrons.

\begin{figure}[htp!]	
\centering
\includegraphics[width=0.48\textwidth,trim=0.0cm 0cm 0cm 0.0cm]{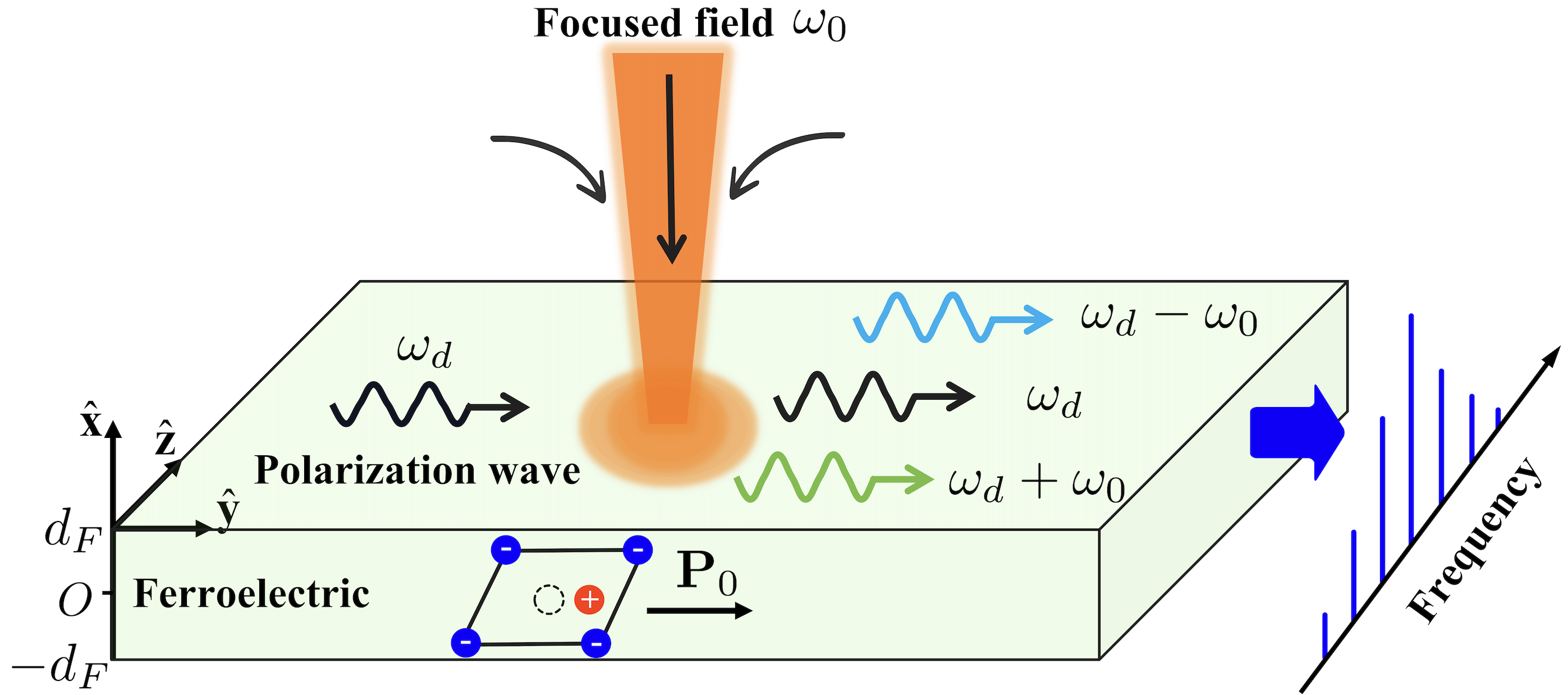}
\caption{Efficient modulation of electric-polarization waves of frequency $\omega_d$ to a frequency comb by a focused electric field of frequency $\omega_0$ in a ferroelectric thin film with a spontaneous polarization ${\bf P}_0$ along the $\hat{\bf y}$-direction. The spacing of the comb frequency components in the comb is $\omega_0$.}
\label{model}
\end{figure}

In this Letter, we demonstrate the dynamic control of anharmonic ferron interactions via photon excitation, based on which we propose the frequency comb of electric-polarization waves in ferroelectrics. As illustrated in Fig.~\ref{model}, driven by a focused optical field of sub-terahertz frequency $\omega_0$, the electric-polarization waves of frequency $\omega_d$ develop many sidebands equally spaced by $\omega_0$ around $\omega_d$, which arises from cascaded sum- and difference-frequency scattering processes triggered by the strong ferroelectric nonlinearity. 
Unlike acoustic phonon and magnon combs, the ferron frequency comb naturally operates in the terahertz regime and derives its strong nonlinearity directly from the intrinsic anharmonicity of the ferroelectric free energy, requiring no nonlinear elastic media. Moreover, the number of comb teeth exhibits a direct correlation with the magnitude of static electric polarization carried by ferrons. This enables the frequency comb to serve as a sensitive nonlinear spectroscopic tool for detecting and characterizing ferrons. By tuning the excitation wavevector, one can perform tomography of the polarization distribution of ferron modes across the Brillouin zone, offering a pathway to visualize the ferrons.

\textit{Nonlinear ferron dynamics}.---We illustrate the principle by considering a thin ferroelectric insulator film located in the $yz$-plane ${\pmb \rho}=y\hat{\bf y}+z\hat{\bf z}$, with a thickness of $2d_F$ along the normal $\hat{\bf x}$-direction (Fig.~\ref{model}). 
The electric polarization ${\bf P}({\bf r},t)={\bf P}_0+\delta{\bf p}({\bf r},t)$, where ${\bf P}_0$ is the spontaneous polarization at equilibrium and $\delta{\bf p}({\bf r},t)$ represents its fluctuation, contributes the free energy $F=\int d{\bf r} {\cal F}({\bf r})$, with the free-energy density~\cite{Zhou2023,Tomeno1988,Scrymgeour2005,Hlinka2006,Chandra_book}  
\begin{align}
    \mathcal{F}= \frac{\alpha_1}{2} P_y^2 + \frac{\alpha_2}{4} P_y^4 +\frac{\alpha_3} {2} (P_x^2 +P_z^2) - \frac{1}{2}{\bf E}_d \cdot \mathbf{P}-{E}_yP_y.
    \label{free_energy}
\end{align}
Here, $\alpha_1<0$ and $\{\alpha_2, \alpha_3\}>0$ are the Landau coefficients, and ${\bf E}_d({\bf P})$ is the electric dipolar field generated by the electric polarization  (refer to the Supplementary Material~(SM) for its calculation~\cite{supplement}).
 The equilibrium polarization ${\bf P}_0=P_{0y}\hat{\bf y}$ is determined by minimizing the free energy, leading to $\alpha_1P_{0y}+\alpha_2P^3_{0y}=E_y$.
$P_{0y}=\pm P_0+{E_y}/({\alpha_1+3\alpha_2P_0^2})$ when $E_y=\pm |E_y|$ are thereby tunable by the applied electric field, in which $P_0=\sqrt{-\alpha_1/\alpha_2}=\{0.746,0.753,0.265\}$~C/m$^{2}$ for LiNbO$_3$, PbTiO$_3$, and BaTiO$_3$ at room temperature~\cite{Tang2022}.

The fluctuation of the electric polarization is governed by the Landau-Khalatnikov-Tani (LKT) equation~\cite{Zhou2023,Rodriguez-Suarez2024,Tani1969,Ishibashi1989,Sivasubramanian2004,Widom2010}
\begin{align}
    m_p\ddot{\bf P}=-\left({\partial F}/{\partial {\bf P}}\right)_T,
    \label{LKT}
\end{align}
where the inertial $m_p=1/(\varepsilon _0 \Omega_p^2)$ is governed by the ionic plasma frequency $\Omega_p$ and the vacuum permittivity $\varepsilon_0$. Such an equation of motion is highly nonlinear since the free-energy density \eqref{free_energy} contains strong anharmonicity. We express $\delta {\bf p}=\delta {\bf p}_l+\delta {\bf p}'$, in which $\delta {\bf p}_l$ accounts for the linear response, while $\delta {\bf p}'$ is responsible for the nonlinearity.

In the linear-response regime, the polarization fluctuation is governed by the linearized LKT equation
\begin{align}
    &(1/\Omega_p^2) \delta\ddot{ {\bf p}}_l(\pmb{\rho},t) + {\cal K} \delta {\bf p}_{l}(\pmb{\rho},t)= \varepsilon_0 {\bf E}_{d}(\delta {\bf p}_l),
    \label{linear_response1}
\end{align}
where ${\cal K}={\rm diag}(K_{\perp},K_{\parallel},K_{\perp})$. The anisotropic ``stiffness" constants $K_{\perp}=\varepsilon_0\alpha_3$ and $K_{\parallel}=\varepsilon_0(\alpha_1+3\alpha_2P_{0y}^2)$ lead to three modes of electric-polarization fluctuation. Mode ``1" with polarization ${\bf e}_{1,{\bf k}}=(1,0,0)^T$ oscillates along the film normal $\hat{\bf x}$-direction and holds an isotropic dispersion $\omega_{1,{\bf k}}=\Omega_p\sqrt{(1-e^{-2d_Fk})/(2d_Fk)+K_{\perp}}$. The other two modes ``$\pm$" oscillate in the film plane with polarizations ${\bf e}_{\pm,{\bf k}}=\sqrt{1/[(\nu_{\bf k}\pm u_{\bf k})^2+1]}(0,\nu_{\bf k}\pm u_{\bf k},1)^T$, which is highly anisotropic with dispersions $\omega_{\pm,{\bf k}}={\Omega_p}\sqrt{\Omega_a({\bf k})\pm \Omega_b({\bf k})}$ (refer to the SM~\cite{supplement}). Here, $\Omega_a({\bf k})=(B_kk^2+K_{\parallel}+K_{\perp})/2$, $\Omega_b({\bf k})=(1/2)\sqrt{[B_k(k_y^2-k_z^2)+K_{\parallel}-K_{\perp}]^2+(2B_kk_yk_z)^2}$, $u_{\bf k}=\Omega_b({\bf k})/({B_{k}k_yk_z})$, and $\nu_{\bf k}=\sqrt{u_{\bf k}^2-1}$, where $B_k=[2d_F-(1-e^{-2d_Fk})/k]/(2d_Fk^2)$.

Disregarding the high-order quartic nonlinearity in the free energy (\ref{free_energy}), we treat the cubic nonlinearity as a perturbation and incorporate it into the equation of motion. In the frequency domain, this leads to 
\begin{align}
    &\begin{pmatrix}
        {\omega^2}/{\Omega_p^2}- K_{\parallel}-B_{k}k^2_y & -B_{k}k_yk_z \\
        -B_{k}k_yk_z  & {\omega^2}/{\Omega_p^2}- K_{\perp}-B_{k}k^2_z 
        \end{pmatrix}
        \begin{pmatrix}
        \delta p'_{y,{\bf k}}(\omega) \\
        \delta p'_{z,{\bf k}}(\omega)
    \end{pmatrix}\nonumber\\
    &=\frac{1}{\sqrt{V_f}}\begin{pmatrix}
    3\alpha_2P_{0y}\varepsilon_0\sum_{\bf k'}f_{{\bf k},{\bf k}'}(\omega)\\
        0
    \end{pmatrix},
    \label{motion1}
\end{align}
in which $f_{{\bf k},{\bf k'}}(\omega)=\int\delta p^l_{y,{\bf k}-{\bf k'}}(t)\delta p^l_{y,{\bf k'}}(t)e^{i\omega t}dt$ and $V_f$ is the crystal volume.
Accordingly, the polarization fluctuation induced by the inharmonic potential energy
\begin{align}
    \delta {\bf p}'_{{\bf k}}(\omega)=\frac{1}{\sqrt{V_f}}\sum_{\bf k'}{\bf F}_{\bf k}(\omega)f_{{\bf k},{\bf k'}}(\omega),
    \label{p_2}
\end{align}
where $f_{{\bf k},{\bf k'}}(\omega)$ serves as the source and 
the vector 
\begin{align}
{\bf F}_{\bf k}(\omega)&=
\begin{pmatrix}
    F_{y,{\bf k}}(\omega)\\
    F_{z,{\bf k}}(\omega)
\end{pmatrix}=\frac{3\alpha_2P_{0y}\varepsilon_0\Omega_p^4}{(\omega^2-\omega^2_{+,{\bf k}})(\omega^2-\omega^2_{-,{\bf k}})}\nonumber\\
&\times
\begin{pmatrix}
    {\omega^2}/{\Omega_p^2}-K_{\perp}-B_{k}k^2_z \\
    B_{k}k_yk_z
\end{pmatrix}
\end{align}
characterizes the system's response to a drive at frequency $\omega$.
As addressed in the denominator of ${\bf F}_{\bf k}(\omega)$, a strong response occurs when the frequency $\omega$ matches one of the mode frequencies $\omega_{\pm,{\bf k}}$.

With these modes, we quantize the electric polarization according to $\delta\hat{\bf p}=\delta\hat{\bf p}_l+\delta\hat{\bf p}'$. 
For the linear order, 
\begin{align}
 \delta \hat{{\bf p}}_{\bf k}^l(t)=\sum_{\lambda}\sqrt{\frac{\hbar}{2m_p\omega_{\lambda,{\bf k}}}}{\bf e}_{\lambda,{\bf k}}\left[\hat{a}_{\lambda,{\bf k}}(t)+\hat{a}^{\dagger}_{\lambda,-{\bf k}}(t)\right],
\label{delta_p0}
\end{align}
in which $\hat{a}_{\lambda,{\bf k}}$ ($\hat{a}^{\dagger}_{\lambda,{\bf k}}$) is the annihilation (creation) operator of ferron in the mode-$\lambda$ and wave vector ${\bf k}$.
Via substitution of Eq.~(\ref{delta_p0}) into (\ref{p_2}) and performing a Fourier transformation to the real space, the nonlinear-order 
$\delta \hat{\bf p}'(\pmb{\rho},t)=(1/\sqrt{V_f})\sum_{\bf k}e^{i{\bf k}\cdot \pmb{\rho}}(1/2\pi)\int \delta \hat{\bf p}'_{\bf k}(\omega)e^{-i\omega t}d\omega$ is expanded in terms of $\{\hat{a}_{\lambda,{\bf k}},\hat{a}^{\dagger}_{\lambda,{\bf k}}\}$: 
\begin{align}
    &\delta\hat{\bf p}'(\pmb{\rho},t)=\frac{1}{{V_f}}\sum_{{\bf k},{\bf k'}}\sum_{\lambda,\lambda'\in \{+,-\}}u^{{\bf k},{\bf k'}}_{\lambda,\lambda'}e^{i{\bf k}\cdot\pmb{\rho}}\nonumber\\
    &\times \left[{\bf F}_{\bf k}(\omega_{\lambda,{\bf k}-{\bf k'}}+\omega_{\lambda',{\bf k'}})\hat{a}_{\lambda,{\bf k}-{\bf k'}}(t)\hat{a}_{\lambda',{\bf k'}}(t)\right.\nonumber\\
    &\left.+{\bf F}_{\bf k}(\omega_{\lambda,{\bf k}-{\bf k'}}-\omega_{\lambda',-{\bf k'}})\hat{a}_{\lambda,{\bf k}-{\bf k'}}(t)\hat{a}^{\dagger}_{\lambda',-{\bf k'}}(t)\right]+{\rm H.c.},
    \label{non_linear_delta_p}
\end{align} 
where $u^{{\bf k},{\bf k'}}_{\lambda,\lambda'}=\hbar/(2m_p)\sqrt{1/(\omega_{\lambda,{\bf k}-{\bf k'}}\omega_{\lambda',{\bf k'}})}e^y_{\lambda,{\bf k}-{\bf k'}}e^y_{\lambda',{\bf k'}}$. Thereby, the nonlinear electric polarization $\delta\hat{\bf p}'(\pmb{\rho},t)$ oscillates with the sum or difference of ferron mode $\lambda\in \{+,-\}$.

In the long wavelength limit ${\bf k}\rightarrow 0$ in $\delta{\bf p}'_{\bf k}$, taking the ensemble average $\langle\cdots\rangle$ over Eq.~\eqref{non_linear_delta_p} yields a macroscopic uniform polarization by the nonlinear fluctuation
\begin{align}
    {\cal P}_y&=\langle \delta \hat{p}'_y(\pmb{\rho},t) \rangle=\frac{1}{V_f}\sum_{\lambda=\pm}\sum_{{\bf k}}\frac{-\hbar}{2m_p}\frac{3\alpha_2P_{0y}}{\alpha_1+3\alpha_2P_{0y}^2}\nonumber\\
    &\times\frac{({e}^y_{\lambda,{\bf k}})^2}{\omega_{\lambda,{\bf k}}}\left[2N(\omega_{\lambda,{\bf k}})+1\right],
    \label{ferron_py1}
\end{align}
a result consistent with the Green function approach (refer to the SM~\cite{supplement}). Here, the Bose-Einstein distribution function $N(\omega_{\lambda,{\bf k}})$ represents the number of ferrons in the mode $\omega_{\lambda,{\bf k}}$. Even at zero temperature, the ferrons carry a static electric polarization due to the zero-energy fluctuation. In the classical limit, $2N(\omega_{\lambda,{\bf k}}) \gg 1$, such that ${\cal P}_y\approx ({1}/{V_f})\sum_{\lambda,{\bf k}}p_{y,\lambda}({\bf k})N(\omega_{\lambda,{\bf k}})$, in which $p_{y,\lambda}({\bf k})=-{\hbar\partial\omega_{\lambda,{\bf k}}}/{\partial E_y}$ is interpreted as the static electric polarization carried by ferron mode $\{\lambda,{\bf k}\}$.

The free energy (\ref{free_energy}) introduces no non-parabolicities to the transverse fluctuations ($\hat{\bf x}$, $\hat{\bf z}$), so the transverse fluctuations do not carry any electric dipole moment along these directions. Indeed, ${\cal P}_{x/z}=\langle \delta \hat{p}'_{x/z}(\pmb{\rho},t) \rangle=0$, thus the ferron carries only an electrical polarization along the saturation polarization $\hat{\bf y}$-direction.

\begin{figure}[htp!]	
\centering
\includegraphics[width=0.48\textwidth,trim=0.0cm 0cm 0cm 0.0cm]{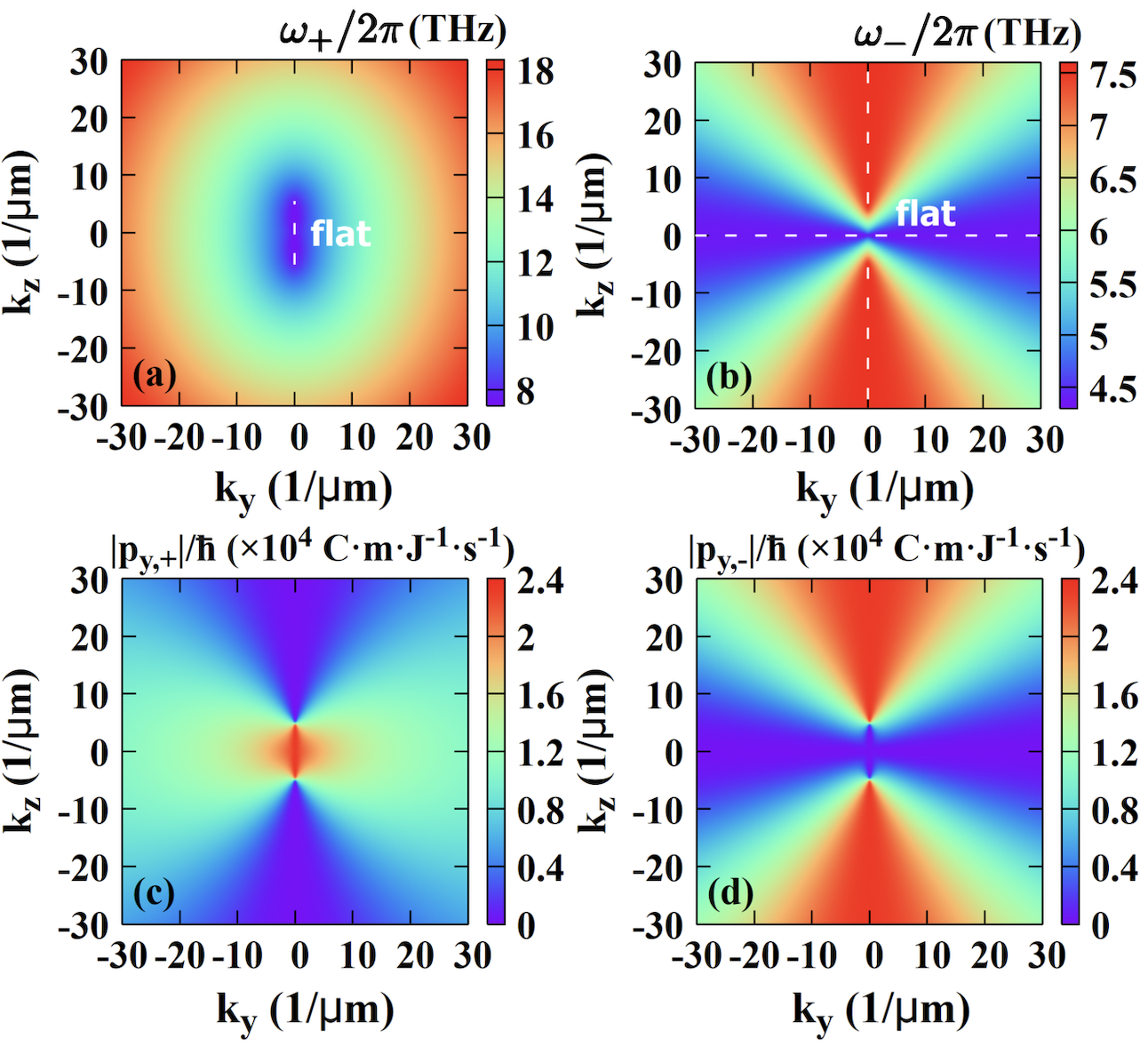}
\caption{Band structures of ferrons of high-frequency $\omega_{+}({\bf k})$ [(a)] and low-frequency $\omega_{-}({\bf k})$ [(b)] modes and their carried static electric polarizations [(c) and (d)].}
\label{band}
\end{figure}

Figure~\ref{band} illustrates the properties of the ``$\pm$" ferron branches for LiNbO$_3$ with thickness $2d_F=10$~nm. For LiNbO$_3$, the plasma frequency $\Omega_p=250$~THz, $\{\alpha_1,\alpha_3\}=\{-2.012,1.345\}\times10^{9}$~${\rm N}\cdot {\rm m}^2/{\rm C}^2$ and $\alpha_2=3.608\times10^{9}$~${\rm N}\cdot {\rm m}^6/{\rm C}^4$ at room temperature~\cite{Scrymgeour2005,Zhou2023} yields the dimensionless stiffness coefficients $\{K_{\perp},K_{\parallel}\}=\{0.012,0.036\}$. Figure~\ref{band}(a) and~\ref{band}(b) show the ferron dispersion relations $\omega_{\pm,{\bf k}}$, which are strongly anisotropic. 
When $k_z\rightarrow 0$, $\omega_{-}\rightarrow \Omega_p\sqrt{K_{\perp}}$ is flat; when $k_y\rightarrow 0$, $\omega_{+}=\Omega_p\sqrt{K_{\parallel}}$ ($\omega_-=\Omega_p\sqrt{K_{\parallel}}$) is flat for $|k_z|\leq (K_{\parallel}-K_{\perp})/d_F$ ($|k_z| > (K_{\parallel}-K_{\perp})/d_F$).
This directional flatness arises from the anisotropic dipolar fields generated by polarization fluctuations along and normal to $P_0\hat{\bf y}$.
Figure~\ref{band}(c) and \ref{band}(d) display the static electric polarization $p_{y,\pm}({\bf k})\propto (e^y_{\pm,{\bf k}})^2/\omega_{\pm,{\bf k}}$ carried by ferrons, where $e^y_{\pm,{\bf k}}$ is the projection of mode's polarization to the saturation $\hat{\bf y}$-direction. We shall demonstrate below that the value of $p_{y,\pm}({\bf k})$ characterizes how the associated ferron modes are effectively modulated by the optical illumination in nonlinear interactions.

\textit{Nonlinear ferron-photon coupling}.---Implementation of the optical field is an excellent approach to excite, detect, and even control the properties of electric polarizations in ferroelectrics~\cite{Choe2025,Zhang2025,Subedi2026}. An optical field
\begin{align}
    {\bf E}_c(\pmb{\rho})=\frac{1}{\sqrt{V_f}}\sum_{\bf k}\pmb{\cal E}_{\bf k}e^{i{\bf k}\cdot\pmb{\rho}}e^{-i\omega_0t}+{\rm H.c.}
\end{align}
of frequency $\omega_0$ and Fourier components $\pmb{\cal E}_{\bf k}$ couples the ferrons by $\hat{H}_c=-\int_{{\bf r}\in V_f}{\bf E}_c(\pmb{\rho})\cdot \delta\hat{\bf p}(\pmb{\rho}) d{\bf r}$.
According to Eq.~(\ref{delta_p0}), the optical field couples to the linear polarization fluctuation $\delta \hat{\bf p}_{l}(\pmb{\rho})$ by
\begin{align}
    \hat{H}_l&=-\sum_{\lambda,{\bf k}}{\bf e}_{\lambda,\bf k}\cdot \left[\pmb{\cal E}^{\ast}_{\bf k}e^{i(\omega_0-\omega_{\lambda,{\bf k}})t}+\pmb{\cal E}_{-\bf k}e^{-i(\omega_0+\omega_{\lambda,{\bf k}})t}\right]\nonumber\\
    &\times\sqrt{\frac{\hbar}{2m_p\omega_{\lambda,{\bf k}}}}\hat{a}_{\lambda,{\bf k}}+{\rm H.c.},
\end{align}
which is responsible for the resonant coherent excitation of ferrons. The nonlinear polarization fluctuation (\ref{non_linear_delta_p}) renders two types of ferron scattering processes by photons, i.e., $\hat{H}_{nl}=\hat{H}_{\rm I}+\hat{H}_{\rm II}$. The first process
\begin{align}
    \hat{H}_{\rm I}&=-\frac{1}{\sqrt{V_f}}\sum_{{\bf k},{\bf k'}}\sum_{\lambda,\lambda'}u^{{\bf k},{\bf k'}}_{\lambda,\lambda'}
    {\bf F}_{\bf k}(\omega_{\lambda,{\bf k}-{\bf k'}}+\omega_{\lambda',{\bf k'}})\nonumber\\
    &\cdot \left[\pmb{\cal E}^{\ast}_{\bf k}e^{i(\omega_0-\omega_{\lambda,{\bf k}-{\bf k'}}-\omega_{\lambda',{\bf k'}})t}+\pmb{\cal E}_{-\bf k}e^{-i(\omega_0+\omega_{\lambda,{\bf k}-{\bf k'}}+\omega_{\lambda',{\bf k'}})t}\right]\nonumber\\
    &\times \hat{a}_{\lambda,{\bf k}-{\bf k'}}\hat{a}_{\lambda',{\bf k'}}+{\rm H.c.}
\end{align}
annihilates or creates two ferrons by the optical field, while the second process
\begin{align}
    \hat{H}_{\rm II}&=-\frac{1}{\sqrt{V_f}}\sum_{{\bf k},{\bf k'}}\sum_{\lambda,\lambda'}u^{{\bf k},-{\bf k'}}_{\lambda,\lambda'}
    {\bf F}_{\bf k}(\omega_{\lambda,{\bf k}+{\bf k'}}-\omega_{\lambda',{\bf k'}})\nonumber\\
    &\cdot \left[\pmb{\cal E}^{\ast}_{\bf k}e^{i(\omega_0-\omega_{\lambda,{\bf k}+{\bf k'}}+\omega_{\lambda',{\bf k'}})t}+\pmb{\cal E}_{-\bf k}e^{-i(\omega_0+\omega_{\lambda,{\bf k}+{\bf k'}}-\omega_{\lambda',{\bf k'}})t}\right]\nonumber\\
    &\times \hat{a}_{\lambda,{\bf k}+{\bf k'}}\hat{a}^{\dagger}_{\lambda',{\bf k'}}+{\rm H.c.}
    \label{H'_c}
\end{align}
scatters ferrons of different wave vectors.

A comparison of these two optical processes is presented in Table~\ref{comparison}.
On one hand, when the optical field is spatially uniform, i.e., ${\bf k}=0$ in its Fourier component $\pmb{\cal E}_{\bf k}$, and its frequency is sufficiently high with $\omega_0\approx \omega_{\lambda,-{\bf k'}}+\omega_{\lambda',{\bf k'}}$, the Hamiltonian $\hat{H}_{\rm I}$ dominates the ferron excitation process, which corresponds to a parametric pumping process, analogous to the magnon excitations~\cite{parametric_pumping1,parametric_pumping2,parametric_pumping3,parametric_pumping4,parametric_pumping5,parametric_pumping6,parametric_pumping7}. 
According to ${\bf F}_{\bf k}(\omega_{\lambda,-{\bf k'}}+\omega_{\lambda',{\bf k'}})=F_{y,{\bf k}}(\omega_{\lambda,-{\bf k'}}+\omega_{\lambda',{\bf k'}})\hat{\bf y}$ when ${\bf k}=0$, an optical field polarized along the static electric polarization $\hat{\bf y}$-direction excites a pair of ferrons in modes $(\lambda,-{\bf k'})$ and $(\lambda',{\bf k'})$. The process can involve either the interband or intraband transitions.
On the other hand, when a spatially inhomogeneous electric field with a lower frequency $\omega_0\approx |\omega_{\lambda',{\bf k'}}-\omega_{\lambda',{\bf k}+{\bf k'}}|$ is applied, $\hat{H}_{\rm II}$ dominates the ferron dynamics, e.g., the ferron frequency comb as we detail below. The frequency comb is produced through cascades of frequency-summation process $\omega_{\lambda',{\bf k}+{\bf k'}}=\omega_{\lambda',{\bf k'}}+\omega_0$ and frequency-difference process
$\omega_{\lambda',{\bf k}+{\bf k'}}=\omega_{\lambda',{\bf k'}}-\omega_0$ in $\hat{H}_{\rm II}$. Such processes occur prominently through the intraband transitions.

\begin{table}
\centering
\caption{Two pathways for optical control of ferron nonlinear dynamics.}
\label{comparison}
\small 
\renewcommand{\arraystretch}{1.0}
\begin{tabular}{
  >{\centering\arraybackslash}m{4.3cm}
  >{\centering\arraybackslash}m{4.3cm}
}
\toprule
\hspace{-0.75cm} \textbf{I: parametric pumping} & \hspace{-0.6cm} \textbf{II: ferron frequency comb} \\
\midrule
\hspace{-0.9cm} \(\hat{H}_{\rm I}\) dominates& 
\hspace{-0.8cm} \(\hat{H}_{\rm II}\) dominates\\
\midrule
\hspace{-0.9cm} \makecell{uniform field 
\((\mathbf{k}=0)\)} & 
\hspace{-0.8cm} \makecell{inhomogeneous field 
\((\mathbf{k}\neq 0)\)} \\
\midrule
\hspace{-0.9cm} \makecell{high frequency:\\
\(\omega_0 \approx \omega_{\lambda,-\mathbf{k}'}+\omega_{\lambda',\mathbf{k}'}\)} & 
\hspace{-0.8cm} \makecell{low frequency:\\
\(\omega_0 \approx \lvert \omega_{\lambda',\mathbf{k}'}-\omega_{\lambda',\mathbf{k}+\mathbf{k}'}\rvert\)} \\
\midrule
\hspace{-0.9cm} field direction $\parallel {\bf P}_0$  & 
\hspace{-0.8cm} field direction $\parallel {\bf P}_0$  \\
\midrule
\hspace{-0.9cm} inter or intraband process& 
\hspace{-0.8cm}intraband process\\
\midrule
\hspace{-0.9cm}\includegraphics[width=1.8cm]{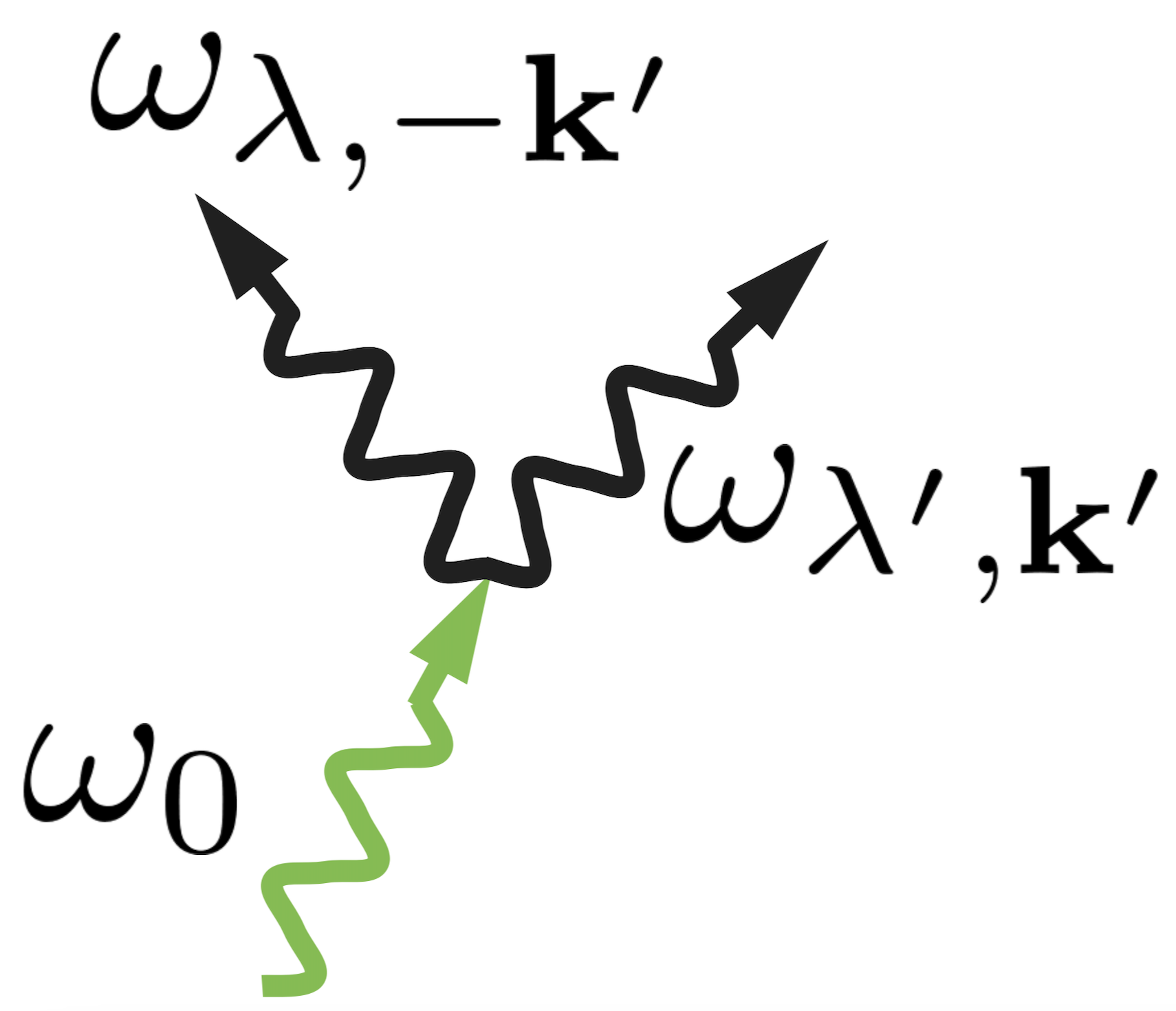} & 
\hspace{-1.0cm}\includegraphics[width=5.2cm]{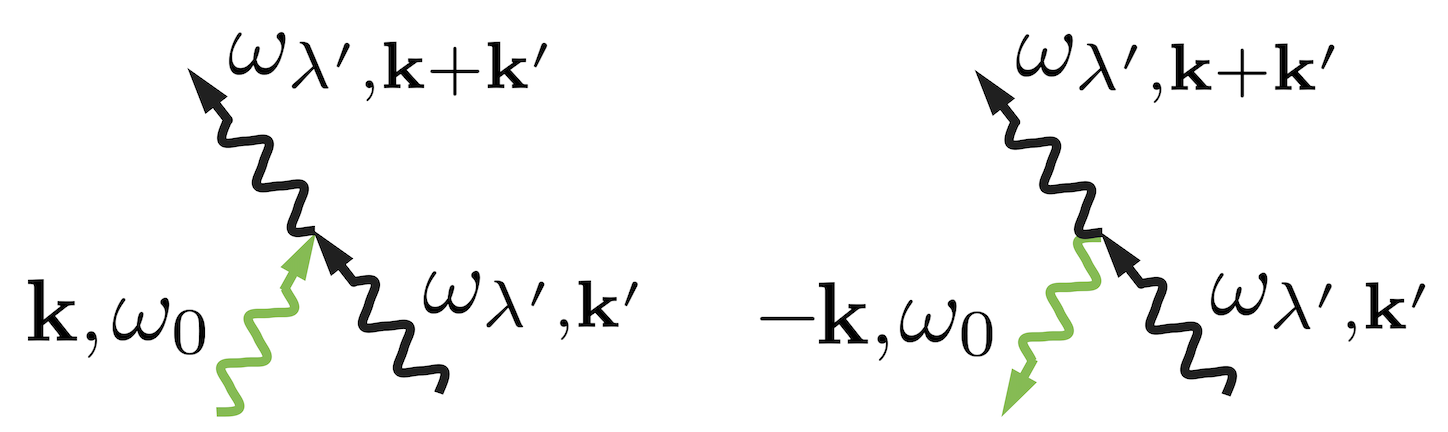} \\
\bottomrule
\end{tabular}
\end{table}

\textit{Ferron frequency comb}.---We now address the nonlinear dynamics of ferron driven by a focused optical field of terahertz frequency $\omega_0 \sim \lvert \omega_{\lambda',\mathbf{k}'}-\omega_{\lambda',\mathbf{k}+\mathbf{k}'}|$, with which $\hat{H}_{\rm II}$ dominates the ferron dynamics. To calculate the modulated mode properties by the focused optical field, we consider the coherent dynamics of ferron mode $({\lambda_0}=\pm,{{\bf k}_0})$ subject to the coherent drive with amplitude $\hbar\Omega_d$ and driven frequency $\omega_d$ chosen to be resonant, i.e., $\omega_d = \omega_{{\lambda_0},{{\bf k}_0}}$. Thus, governed by the effective Hamiltonian $\hat{H}=\sum_{\lambda=\pm,{\bf k}}\hbar\omega_{\lambda,\bf k}\hat{a}^{\dagger}_{\lambda,{\bf k}}\hat{a}_{\lambda,{\bf k}}+\hat{H}_{\rm II}+\hbar\Omega_d(\hat{a}_{{\lambda_0},{{\bf k}_0}}e^{i\omega_dt}+{\rm H.c.})$, the amplitude $\beta_{{\lambda}_0,\mathbf{k}}$ in 
$\langle \hat{a}_{{\lambda}_0,\mathbf{k}} \rangle = \beta_{{\lambda}_0,\mathbf{k}} e^{-i\omega_dt}$ obeys the equation of motion 
\begin{align}
    &\partial_t \beta_{{\lambda}_0,\mathbf{k}}=-[i(\omega_{{\lambda}_0,{\bf k}}-\omega_d)+\alpha\omega_{{\lambda}_0,{\bf k}}]\beta_{{\lambda}_0,\mathbf{k}}-i\Omega_d\delta_{{\bf k},{{\bf k}}_0}\nonumber\\
    &+\frac{i}{\hbar}\sum_{{\bf k}'}g(\lambda_0,{\bf k},{\bf k}')({E}^{\ast}_{y,{\bf k}'}e^{i\omega_0t}+{E}_{y,-{\bf k}'}e^{-i\omega_0t})\beta_{{\lambda}_0,{\bf k}'+{\bf k}},
    \label{beta}
\end{align}
where $E_{y,\mathbf{k}'}={\cal E}_{y,\mathbf{k}'}/\sqrt{V_f}$ and $\alpha=10^{-3}$ phenomenologically accounts for the damping of ferron~\cite{Zhou2023,Choe2025,Zhang2025,Subedi2026,Shen2025}.
Here, the coupling coefficient $g(\lambda_0,{\bf k},{\bf k}')=2u^{{\bf k}',-{\bf k}}_{\lambda_0,\lambda_0}F_{y,{\bf k}'}(\omega_{\lambda_0,{\bf k}'+{\bf k}}-\omega_{\lambda_0,{\bf k}})$
mixes different ferron modes.
The coherent mode $({\lambda_0},{\bf k}_0)$ generates new frequency components through sum- and difference-frequency processes with $\omega_0$. These new components mix further, creating a cascade that populates a series of equally spaced ferron frequency combs.

To be specific, we consider a focused electric field ${\bf E}({\bm \rho})=E_0\Theta(\rho_0-\rho)e^{-i\omega_0t}\hat{\bf y}+{\rm H.c.}$ of amplitude $E_0$, radius $\rho_0$, and frequency $\omega_0$. Its Fourier component $E_{y,{\bf k}'}=(1/A)(2\pi E_0\rho_0/k')[\sin(k'\rho_0)/(k'\rho_0)^2-\cos(k'\rho_0)/(k'\rho_0)]$ reaches its maximum $E_{y,{\bf k'}\rightarrow0}=(1/A)\pi E_0\rho_0^2$ as ${\bf k}'\rightarrow 0$ and decays rapidly as $k'$ increases, where $A$ is the film area. The effective range of $k'$ is $k'\in [-2\pi/\rho_0,2\pi/\rho_0]$, such that $|{\bf k}'|\ll |{\bf k}|$, rendering
\begin{align}
g(\lambda_0,{\bf k},{\bf k}')\approx p_{y,{\lambda_0}}({\bf k})
\end{align}
solely governed by the ferron's static electric polarization.

When ${\bf k}={\bf k}_0$, the nonlinear ferron-photon interaction in Eq.~(\ref{beta}) enables the scattering of ferron of mode $(\lambda_0,{\bf k}_0)$ to the other modes, with efficiency governed by the ferron electric polarization $p_{y,{\lambda_0}}({\bf k}_0)$. The absorption of photon $\omega_0$ excites the ferron of wave vector ${\bf k}$ and generates the first sidebands at frequency $\omega_{{\bf k}}=\omega_{{\bf k}_0}\pm \omega_0$. These sidebands act as new sources to generate the high-order sidebands, with the capability governed by their respective static polarization $p_{y,{\lambda_0}}({\bf k})$. 

\begin{figure}[htp!]	
\centering
\includegraphics[width=0.47\textwidth,trim=0.0cm 0cm 0cm 0.0cm]{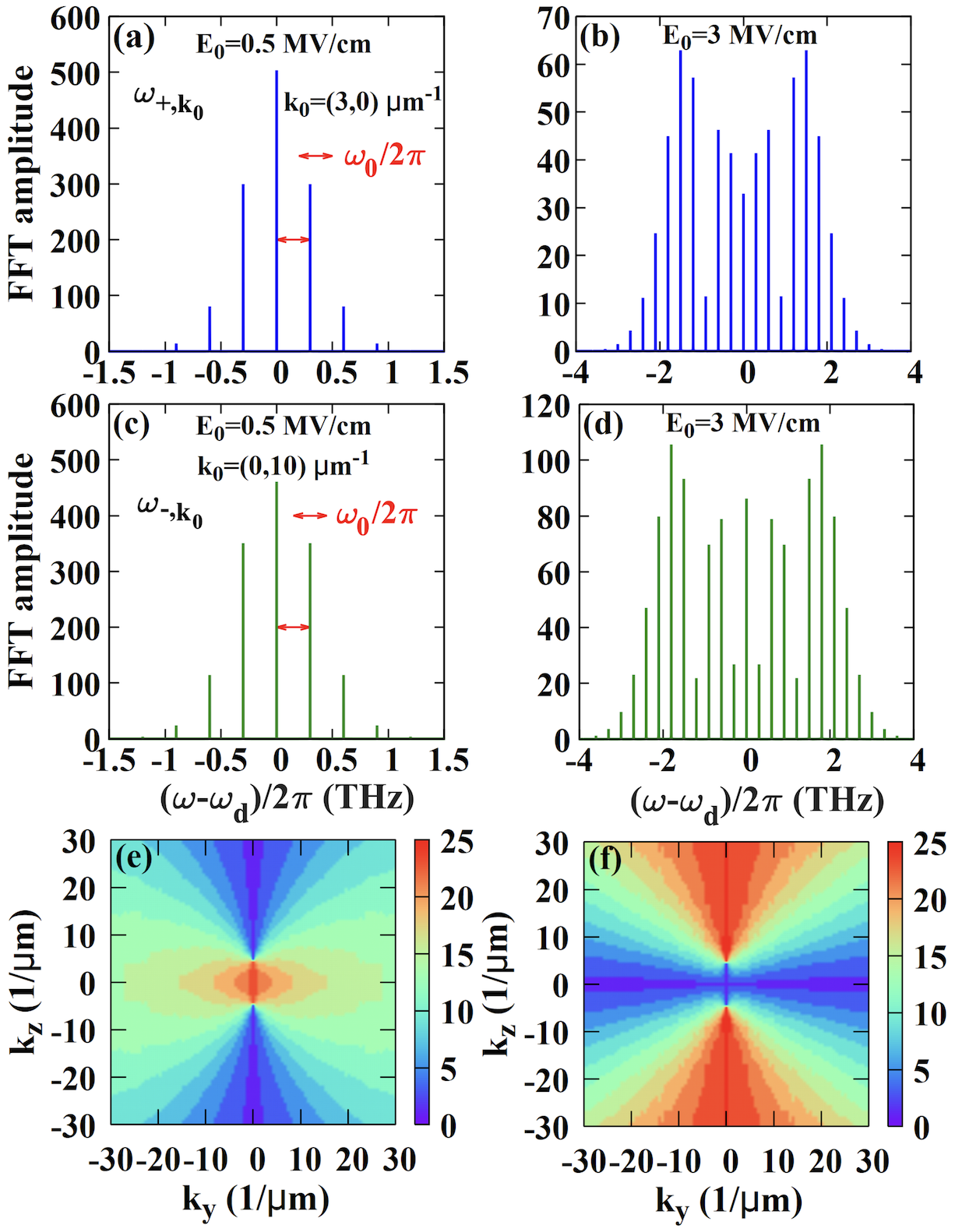}
\caption{Frequency comb spectra of coherent ferrons. (a) and (b):  frequency comb of mode $\{\lambda_0=+,{\bf k}_0=(3,0)~\mu{\rm m}^{-1}\}$ modulated by electric fields of $\{0.5,3\}$~MV/cm. (c) and (d): frequency comb of mode $\{\lambda_0=-,{\bf k}_0=(0,10)~\mu {\rm m}^{-1}\}$. (e) and (f): tomography of ferron electric polarization by the frequency comb teeth number as a function of mode wavevector ${\bf k}$ for the $\omega_+$ [(e)] branch and $\omega_-$ [(f)] ferron branch.}
\label{comb}
\end{figure}

Figure~\ref{comb}(a) and (b) illustrate the frequency comb spectra of the mode $(\lambda_0=+,{\bf k}_0=(3,0)~\mu {\rm m}^{-1})$, generated by the coherent drive with coupling strength $\Omega_d=50$~THz and a focused electric field of radius $\rho_0=500~\mu$m and frequency $\omega_0=1.9$~THz with amplitude strengths $0.5$~MV/cm [(a)] and $3$~MV/cm [(b)], respectively~\cite{E0,Subedi2026}. Such focused fields are feasible in terahertz pump–optical probe experiments and have been implemented to resonantly excite ferrons in van der Waals ferroelectrics~\cite{Subedi2026}.
Figure~\ref{comb}(c) and (d) show the corresponding spectra of the mode  $(\lambda_0=-,{\bf k}_0=(0,10)~\mu {\rm m}^{-1})$, driven by the same field strengths. Due to the Floquet modulation introduced by the optical field at frequency $\omega_0$, the pump mode develops a frequency comb structure with peaks at $\omega_d\pm m\omega_0$, holding very narrow broadening, where $m$ are integers. As the electric field increases, the nonlinear effects become significantly stronger, transferring energy from the pump mode to more sidebands. Indeed, the number of observable comb teeth increases significantly to about 30 as the electric field amplitude $E_0$ increases to $3$~MV/cm [Fig.~\ref{comb}(b) and (d)].

To further trace the origin of the frequency combs, we plot in Fig.~\ref{comb}(e) and (f) the number of comb teeth as a function of the excitation wavevector ${\bf k}$ for the two ferron branches, calculated at $E_0=3$~MV/cm. The results demonstrate the efficiency of the frequency comb is governed by the static electric polarization $p_{y}({\bf k})$ carried by the ferron: the distribution of the teeth number in Fig.~\ref{comb}(e) and (f) shows exactly the same pattern as that of $p_y({\bf k})$ in Fig.~\ref{band}(c) and \ref{band}(d). A pump mode located in a region of large 
$p_y({\bf k})$ interacts more strongly with photons and can efficiently cascade more energy to generate a broad-spectrum comb. In contrast, a small $p_y$ limits the influence of the pump, yielding only a few teeth.

\textit{Conclusion and discussion}.---In conclusion, we develop a quantum theory to account for the nonlinear interaction between ferron and photon by revealing two types of nonlinear optical processes, i.e., the parametric pumping of ferron and the ferron frequency-comb generation. This quantum framework allows the series expansion of the electric-polarization operators in terms of ferron operators, which can readily be extended to other hybridizations, \textit{e.g.}, ferron-phonon and ferron-electron nonlinear couplings. We show that the optical control of electric-polarization nonlinear dynamics is efficient in ferroelectrics: using the existing focused electric field of frequency $\sim{\rm THz}$, the number of teeth spaced by the field frequency in the comb achieves more than thirty. In the long-wavelength limit, the nonlinear ferron-photon coupling strength is reduced exactly to the electric dipole moment carried by the ferron mode, so the ferron frequency comb can realize the tomography of the ferron's electric polarization across the Brillouin zone, which thereby provides strong evidence for the ferrons. Such a frequency comb should inspire useful functionalities in the terahertz regime.

\begin{acknowledgments}
This work is financially supported by the National Natural Science Foundation of China under Grant No.~12374109 and the National Key Research and Development Program of China under Grant No.~2023YFA1406600. 
\end{acknowledgments}


\begin{thebibliography}{99}

\bibitem{optical_comb9} J. N. Eckstein, A. I. Ferguson, and T. W. H{\"a}nsch, High-Resolution Two-Photon Spectroscopy with Picosecond Light Pulses, Phys. Rev. Lett. \textbf{40}, 847 (1978).


\bibitem{optical_comb1} D. J. Jones, S. A. Diddams, J. K. Ranka, A. Stentz, R. S. Windeler, J. L. Hall, and S. T. Cundiff, Carrier-Envelope Phase Control of Femtosecond Mode-Locked Lasers and Direct Optical Frequency Synthesis, Science \textbf{288}, 635 (2000).




\bibitem{optical_comb2} S. T. Cundiff and J. Ye, Colloquium: Femtosecond optical frequency combs, Rev. Mod. Phys. \textbf{75}, 325 (2003).


\bibitem{optical_comb8} A. Dutt, C. Joshi, X. Ji, J. Cardenas, Y. Okawachi, K. Luke, A. L. Gaeta, and M. Lipson, On-chip dual-comb source for spectroscopy, Sci. Adv. \textbf{4}, e1701858 (2018).


\bibitem{optical_comb6} A. Pasquazi, M. Peccianti, L. Razzari, D. J. Moss, S. Coen, M. Erkintalo, Y. K. Chembo, T. Hansson, S. Wabnitz, P. De{\'l}Haye, X. Xue, A. M. Weiner, and R. Morandotti, Micro-combs: A novel generation of optical sources, Phys. Rep. \textbf{729}, 1 (2018).


\bibitem{optical_comb3} N. Picqu{\'e} and T. W. H{\"a}nsch, Frequency comb spectroscopy, Nat. Photon. \textbf{13}, 146 (2019).



\bibitem{optical_comb5} T. Fortier and E. Baumann, 20 years of developments in optical frequency comb technology and applications, Commun. Phys. \textbf{2}, 153 (2019).



\bibitem{optical_comb7} M. G. Suh, X. Yi, Y. H. Lai, S. Leifer, I. S. Grudinin, G. Vasisht, E. C. Martin, M. P. Fitzgerald, G. Doppmann, J. Wang, D. Mawet, S. B. Papp, S. A. Diddams, C. Beichman, and K. Vahala, Searching for exoplanets using a microresonator astrocomb, Nat. Photon. \textbf{13}, 25 (2019).


\bibitem{optical_comb4} H. Zhang, B. Chang, Z. Li, Y.-P. Liang, C.-Y. Qin, C. Wang, H.-D. Xia, T. Tan, and B.-C. Yao, Coherent optical frequency combs: From principles to applications, Journal of Electronic Science and Technology \textbf{20}, 100157 (2022).


\bibitem{phonon1} L. S. Cao, D. X. Qi, R. W. Peng, M. Wang, and P. Schmelcher, Phononic Frequency Combs through Nonlinear Resonances, Phys. Rev. Lett. \textbf{112}, 075505 (2014).



\bibitem{phonon2} A. Ganesan, C. Do, and A. Seshia, Phononic Frequency Comb Via Intrinsic Three-Wave Mixing, Phys. Rev. Lett. \textbf{118}, 033903 (2017).



\bibitem{phonon3} A. Ganesan, C. Do, and A. Seshia, Excitation of coupled phononic frequency combs via two-mode parametric three-wave mixing, Phys. Rev. B \textbf{97}, 014302 (2018).



\bibitem{phonon4} J. Zhang, B. Peng, S. Kim, F. Monifi, X. Jiang, Y. Li, P. Yu, L. Liu, Y. X. Liu, A. Alu, and L. Yang, Optomechanical dissipative solitons, Nature (London) \textbf{600}, 75 (2021).



\bibitem{phonon5} S. Wu, Y. Liu, Q. Liu, S. P. Wang, Z. Chen, and T. Li, Hybridized frequency combs in multimode cavity electromechanical system, Phys. Rev. Lett. \textbf{128}, 153901 (2022).



\bibitem{phonon6} M. H. de Jong, A. Ganesan, A. Cupertino, and R. A. Norte, Mechanical overtone frequency combs, Nat. Commun. \textbf{14}, 1458 (2023).


\bibitem{phonon7} J. Sun, S. Yu, H. Zhang, D. Chen, X. Zhou, C. Zhao, D. D. Gerrard, R. Kwon, G. Vukasin, D. Xiao, T. W. Kenny, X. Wu, and A. Seshia, Generation and Evolution of Phononic Frequency Combs via Coherent Energy Transfer between Mechanical Modes, Phys. Rev. Applied. \textbf{19}, 014031 (2023).


\bibitem{phonon10} C. Cai, X.-H. Zhou, W. Yu, and T. Yu, Acoustic frequency multiplication and pure second-harmonic generation of phonons by magnetic transducers, Phys. Rev. B \textbf{107}, L100410 (2023).



\bibitem{phonon8} Y. Wang, M. Zhang, Z. Shen, G.-T. Xu, R. Niu, F.-W. Sun, G.-C. Guo, and C.-H. Dong, Optomechanical Frequency Comb Based on Multiple Nonlinear Dynamics, Phys. Rev. Lett.\textbf{132}, 163603 (2024).



\bibitem{phonon9} Z. Yu, Z. Jin, Q. Zheng, and P. Yan, Magnon-Driven Phononic Frequency Comb in Linear Elastic Media, arXiv:2505.19673.



\bibitem{magnon1} Z. Wang, H. Y. Yuan, Y. Cao, Z.-X. Li, R. A. Duine, and P. Yan, Magnonic Frequency Comb through Nonlinear Magnon-Skyrmion Scattering, Phys. Rev. Lett. \textbf{127}, 037202 (2021).



\bibitem{magnon2} T. Hula, K. Schultheiss, F. J. T. Gon\c{c}alves, L. K{\"o}rber, M. Bejarano, M. Copus, L. Flacke, L. Liensberger, A. Buzdakov, A. K{\'a}kay, M. Weiler, R. Camley, J. Fassbender, and H. Schultheiss, Spin-wave frequency combs, Appl. Phys. Lett. \textbf{121}, 112404 (2022).  



\bibitem{magnon3} Z. Wang, H. Y. Yuan, Y. Cao, and P. Yan, Twisted Magnon Frequency Comb and Penrose Superradiance, Phys. Rev. Lett. \textbf{129}, 107203 (2022).


\bibitem{magnon6} H. Xiong, Magnonic frequency combs based on the resonantly enhanced magnetostrictive effect, Fundamental Research \textbf{3}, 8 (2023).


\bibitem{magnon4} G.-T. Xu, M. Zhang, Y. Wang, Z. Shen, G.-C. Guo, and C.-H. Dong, Magnonic Frequency Comb in the Magnomechanical Resonator, Phys. Rev. Lett. \textbf{131}, 243601 (2023).



\bibitem{magnon8} J. W. Rao, B. Yao, C. Y. Wang, C. Zhang, T. Yu, and W. Lu, Unveiling a Pump-Induced Magnon Mode via Its Strong Interaction with Walker Modes, Phys. Rev. Lett. \textbf{130}, 046705 (2023).



\bibitem{magnon9} C. Wang, J. Rao, Z. Chen, K. Zhao, L. Sun, B. Yao, T. Yu, Y.-P. Wang, and W. Lu, Enhancement of magnonic frequency combs by exceptional points, Nat. Phys. \textbf{20}, 1139 (2024).



\bibitem{magnon5} G.-T. Xu, Z. Shen, M. Zhang, Y. Wang, S. Wan, Y. Yang, T. Zhang, L. Bi, F.-W. Sun, G.-C. Guo, and C.-H. Dong, Kerr-Induced Synchronization of a Broadband Magnon-Phonon Hybrid Frequency Comb, Phys. Rev. Lett. \textbf{135}, 203604 (2025).


\bibitem{magnon7} G. Lan, K.-Y. Liu, Z. Wang, F. Xia, H. Xu, T. Guo, Y. Zhang, B. He, J. Li, C. Wan, G. E. W. Bauer, P. Yan, G.-Q. Liu, X.-Y. Pan, X. Han, and G. Yu, Coherent harmonic generation of magnons in spin textures, Nat. Commun. \textbf{16}, 1178 (2025).




\bibitem{Bauer2021} G. E. W. Bauer, R. Iguchi, and K.-i. Uchida, Theory of Transport in Ferroelectric Capacitors, Phys. Rev. Lett. \textbf{126}, 187603 (2021).



\bibitem{Tang2022_thermoelectric} P. Tang, R. Iguchi, K.-i. Uchida, and G. E. W. Bauer, Thermoelectric polarization transport in ferroelectric ballistic point contacts, Phys. Rev. Lett. \textbf{128}, 047601 (2022).


\bibitem{Tang2022} P. Tang, R. Iguchi, K.-i. Uchida, and G. E. W. Bauer, Excitations of the ferroelectric order, Phys. Rev. B \textbf{106}, L081105 (2022).

\bibitem{Hu2022} S. Zhuang and J.-M. Hu, Role of polarization-photon coupling in ultrafast terahertz excitation of ferroelectrics, Phys. Rev B \textbf{106}, L140302 (2022).


\bibitem{Zhou2023} X.-H. Zhou, C. Cai, P. Tang, R. L. Rodr\'iguez-Su\'arez, S. M. Rezende, G. E. W. Bauer, and T. Yu, Surface Ferron Excitations in Ferroelectrics and Their Directional Routing, Chin. Phys. Lett. \textbf{40}, 087103 (2023).



\bibitem{Tang2024} P. Tang and G. E. W. Bauer, Electric analog of magnons in order-disorder ferroelectrics, Phys. Rev. B \textbf{109}, L060301 (2024).


\bibitem{Rodriguez-Suarez2024} R. L. Rodr{\'i}guez-Su{\'a}rez, X.-H. Zhou, C. Y. Cai, P. Tang,
T. Yu, G. E. W. Bauer, and S. M. Rezende, Surface and volume modes of polarization waves in ferroelectric films, Phys. Rev. B \textbf{109}, 134307 (2024).


\bibitem{Zhu2024} Y. Zhu, T. Chen, A. Ross, B. Wang, X. Guo, V. Gopalan, L.-Q. Chen, and J.-M. Hu, Theory of nonlinear terahertz susceptibility in ferroelectrics, Phys. Rev. B \textbf{110}, 054311 (2024).

\bibitem{Morozovska2025} A. N. Morozovska, E. A. Eliseev, O. V. Bereznikov, M. Ye. Yelisieiev, G.-D. Zhao, Y. Zhu, V. Gopalan, L.-Q. Chen, J.-M. Hu, and Y. M. Vysochanskii, Flexocoupling-induced phonons and ferrons in van der Waals ferroelectrics, Phys. Rev. B \textbf{112}, 014110 (2025).


\bibitem{Bauer2022} G. E. W. Bauer, P. Tang, R. Iguchi, and K.-i. Uchida, Magnonics vs. Ferronics, Journal of Magnetism and Magnetic Materials \textbf{541}, 168468 (2022).

\bibitem{Bauer2023} G. E. W. Bauer, P. Tang, R. Iguchi, J. Xiao, K. Shen, Z. Zhong, T. Yu, S.M. Rezende, J.P. Heremans, and K. Uchida, Polarization transport in ferroelectrics, Phys. Rev. Applied. \textbf{20}, 050501 (2023).


\bibitem{Yu2026_review} T. Yu, X.-H. Zhou, G. E. W. Bauer, and I. Bobkova, Electromagnetic proximity effects at heterointerfaces, Physics Reports \textbf{1151}, 1 (2026).




\bibitem{Kranendonk1958} J. Van Kranendonk and J. H. Van Vleck, Spin waves, Rev. Mod. Phys. \textbf{30}, 1 (1958).


\bibitem{Stancil_book} D. D. Stancil and A. Prabhakar, \textit{Spin Waves: Theory and Applications}, 1st ed. (Springer New York, NY, 2009).


\bibitem{Chumak2015} A. V. Chumak, V. I. Vasyuchka, A. A. Serga, and B. Hillebrands, Magnon spintronics, Nature Physics \textbf{11}, 453 (2015).


\bibitem{Tang2023} P. Tang, K.-i. Uchida, and G. E. W. Bauer, Nonlocal drag thermoelectricity generated by ferroelectric van der waals heterostructures, Phys. Rev. B \textbf{107}, L121406 (2023).


\bibitem{Nursagatov2026} M. Nursagatov, X. Ye, G. A. Bobkov, T. Yu, and I. V. Bobkova, Ferron-Polaritons in Superconductor/Ferroelectric/Superconductor Heterostructures, arXiv:2602.05473.




\bibitem{Choe2025} J. Choe, T. Handa, C.-Y. Huang, A. K. Liston, J. Cox, J. Stensberg, Y. Hong, D. G. Chica, D. Xu, F. Tay, S. Husremovic, V. da Silveira Lanza Avelar, E. A. Arsenault, Z. Zhang, J. McIver, D. N. Basov, M. Delor, X. Roy, and X.-Y. Zhu, Observation of coherent ferrons, arXiv:2505.22559.



\bibitem{Zhang2025} B. Zhang, R. Duan, S. S. Mishra, S. Jana, J. Kim, T. T. Caiwei, Y. J. Tan, W. Wang, P. T. C. Ietro, Z. Liu, and R. Singh, Electric-field control of giant ferronics, arXiv:2509.06057.



\bibitem{Subedi2026} S. Subedi, W. Fang, F. Fei, Z. Zhai, J. P. Rollins, C. Fox, A. Drew, B. Lv, Y. Ping, and J. Xiao, Electrically switchable ferron upconversion in a van der Waals ferroelectric, arXiv:2603.19394.


\bibitem{Shen2025} K. Shen, P. Tang, X. Chen, Y. Gao, Y. Fan, Z. Guo, Y. Wei, H. Jiang, X. Zhang, M. Wang, P. He, W. Shi, J. Han, Y. Wu, J. Shen, Q. Liu, G. E. W. Bauer, and M. Liu, Observation of ferron transport in ferroelectrics, arXiv:2505.24419.


\bibitem{Wooten2023} B. L. Wooten, R. Iguchi, P. Tang, J. S. Kang, K. i. Uchida, G. E. W. Bauer, and J. P. Heremans, Electric field–dependent phonon spectrum and heat conduction in ferroelectrics, Science Advances \textbf{9}, eadd7194 (2023).





\bibitem{Tomeno1988} I. Tomeno and S. Matsumura, Dielectric properties of LiTaO$_3$, Phys. Rev. B \textbf{38}, 606 (1988).


\bibitem{Scrymgeour2005} D. A. Scrymgeour, V. Gopalan, A. Itagi, A. Saxena, and P. J. Swart, Phenomenological theory of a single domain wall in uniaxial trigonal ferroelectrics: Lithium niobate and lithium tantalate, Phys. Rev. B \textbf{71}, 184110 (2005).


\bibitem{Hlinka2006} J. Hlinka and P. M{\'a}rton, Phenomenological model of a $90^{\circ}$ domain wall in BaTiO$_3$-type ferroelectrics, Phys. Rev. B \textbf{74}, 104104 (2006).

\bibitem{Chandra_book} P. Chandra and P. B. Littlewood, \textit{A Landau Primer for Ferroelectrics in Physics of Ferroelectrics}, pp 69–116, (Springer, Berlin, 2007).

\bibitem{supplement} See Supplemental Material at [...] for the derivation of ferron modes, the quantization of the electric fluctuation, and the calculation of the electric polarization carried by ferrons. 


\bibitem{Tani1969} K. Tani, Dynamics of Displacive-Type Ferroelectrics –Soft Modes–, J. Phys. Soc. Jpn. \textbf{26}, 93 (1969).


\bibitem{Ishibashi1989} Y. Ishibashi, Phenomenological theory of domain walls, Ferroelectrics \textbf{98}, 193 (1989).


\bibitem{Sivasubramanian2004} S. Sivasubramanian, A. Widom, and Y. N. Srivastava, Physical Kinetics of Ferroelectric Hysteresis, Ferroelectrics \textbf{300}, 43 (2004).


\bibitem{Widom2010} A. Widom, S. Sivasubramanian, C. Vittoria, S. Yoon, and Y. N. Srivastava, Resonance damping in ferromagnets and ferroelectrics, Phys. Rev. B \textbf{81}, 212402 (2010).




\bibitem{parametric_pumping1} N. Bloembergen and R. W. Damon, Relaxation effects in ferromagnetic resonance, Phys. Rev. \textbf{85}, 699 (1952).


\bibitem{parametric_pumping2} N. Bloembergen and S. Wang, Relaxation effects in para- and ferromagnetic
resonance, Phys. Rev. \textbf{93}, 72 (1954).


\bibitem{parametric_pumping3} P. W. Anderson and H. Suhl, Instability in the motion of ferromagnets at
high microwave power levels, Phys. Rev. \textbf{100}, 1788 (1955).



\bibitem{parametric_pumping4} R. Verba, V. Tiberkevich, I. Krivorotov, and A. Slavin, Parametric excitation
of spin waves by voltage-controlled magnetic anisotropy, Phys. Rev. Appl. \textbf{1}, 044006 (2014).


\bibitem{parametric_pumping5} Y.-J. Chen, H. K. Lee, R. Verba, J. A. Katine, I. Barsukov, V. Tiberkevich,
J. Q. Xiao, A. N. Slavin, and I. N. Krivorotov, Parametric resonance of magnetization excited by electric field, Nano Lett. \textbf{17}, 572 (2017).


\bibitem{parametric_pumping6} T. Br{\"a}cher, P. Pirro, and B. Hillebrands, Parallel pumping for magnon spintronics: Amplification and manipulation of magnon spin currents on the micron-scale, Phys. Rep. \textbf{699}, 1 (2017).


\bibitem{parametric_pumping7} G. Okano and Y. Nozaki, Spin waves parametrically excited via three-magnon
scattering in narrow NiFe strips, Phys. Rev. B \textbf{100}, 104424 (2019).


\bibitem{E0} H. Hirori, A. Doi, F. Blanchard, and K. Tanaka, Single-cycle terahertz pulses with amplitudes exceeding 1 MV/cm generated by optical rectification in LiNbO$_3$, Appl. Phys. Lett. \textbf{98}, 091106 (2011).


\end{thebibliography}
\end{document}